\documentclass[conference]{IEEEtran}

\usepackage{amsmath}
\usepackage{amssymb}
\usepackage[dvips]{graphicx}
\usepackage{epsfig}

\newtheorem{definition}{Definition}[section]
\newtheorem{theorem}{Theorem}[section]

\begin{document}

\title{Strong Analytic Controllability for Hydrogen Control Systems}

\author{\authorblockN{Chunhua Lan}
\authorblockA{Department of Electrical \\ and Systems Engineering\\
Washington University\\
St. Louis, MO 63130-4899 \\
Email: lancb@bc.edu} \and
\authorblockN{Tzyh-Jong Tarn }
\authorblockA{Department of Electrical \\ and Systems Engineering\\
Washington University\\
St. Louis, MO 63130-4899 \\
Email: tarn@wuauto.wustl.edu}
\and
\authorblockN{Quo-Shin Chi }
\authorblockA{Department of Mathematics \\
Washington University\\
St. Louis, MO 63130-4899 \\
Email: chi@math.wustl.edu}
\and
\authorblockN{John W. Clark}
\authorblockA{Department of Physics \\
Washington University\\
St. Louis, MO 63130-4899\\
Email:jwc@wuphys.wustl.edu }}

\maketitle

\begin{abstract}
 The realization and representation
of $so(4,2)$ associated with the hydrogen atom Hamiltonian are
derived. By choosing operators from the realization of $so(4,2)$
as interacting Hamiltonians, a hydrogen atom control system is
constructed, and it is proved that this control system is strongly
analytically controllable based on a time-dependent strong
analytic controllability theorem.
\end{abstract}

\IEEEpeerreviewmaketitle

\section{Introduction}

In the microscopic world ruled by quantum mechanics, most
interesting phenomena involve change of continuous quantum
operators acting on infinite-dimensional state spaces. For
instance, developments in quantum error correction \cite {Lloyd_4,
Braunstein_1, Braunstein_2} and quantum teleportation \cite
{Braunstein_3, Furusawa} require addressing the question of
continuous variables. It is noted that real world quantum systems
are influenced by interaction with their changing environment to a
greater or lesser extent. The structural Hamiltonians used to
describe these open quantum systems are time-dependent ones, as is
the case in \cite {Thorwart, Viola}. New research pays more and
more attention in manipulation of time-dependent quantum systems
acting on infinite-dimensional state spaces, which contrasts to
the manipulation of discrete systems with finite-dimensional state
spaces in traditional quantum computation.

From a theoretical point of view, the essential problem of the
manipulation of quantum dynamical phenomena is the problem of
control. For instance, the field of nuclear magnetic resonance is
largely concerned with the geometric control of collections of
interacting nuclear spins \cite {Lloyd_1, Slichter, Ernst_1,
Elmsley}. The methods used to derive these results are those of
geometric control applied to quantum systems: these
group-theoretic methods allow easy mathematical treatment of
Hamiltonian systems.  Not only do certain systems possess obvious
symmetries in their spectra which allows the application of group
theory , but it has also been suggested that group theory might
help to look for certain transformations which allow passing from
one level state to another, and thus get new insight into the
structure of the system \cite {Dothan_2}.

In this paper, we use group-theoretic methods to study the control
problem of a time-dependent hydrogen atom system acting on
infinite-dimensional state spaces. As the simplest of all atoms,
hydrogen atom plays a central role in quantum mechanical theory.
It is the only atom for which the energy eigenvalue problem has
been solved exactly. Wave functions of the hydrogen atom are the
starting point for the description of all atoms and molecules.
Understanding how to control the hydrogen atoms is the starting
point for understanding on how to control more complicated atoms
and molecules. From the design of a control for atoms and
molecules via perturbation/approximation method, the solution of
the control problem of hydrogen atom furnishes the first step.
From energy spectrum point of view, hydrogen atom possesses both
discrete (bound state) and continuous (scattering states) spectra.
Understanding the control problem of hydrogen atom will help us to
understand the controlled transition from bound states to
scattering states and vise versa.

 It is well known that the bound state
subspace of a hydrogen atom
 for a fixed energy $E_n$ spans a representation space of
 dimension $n^2$ of the symmetry group $SO(4)$ \cite {Barut_1}. The symmetry group is $SO(3,1)$
 for scattering states.  When the atom interacts with an external
 electromagnetic field, it can make transitions to other energy subspaces, because
 energy is transferred to or from the
 atom.  Thus we can connect a representation space of $SO(4)$ or
 $SO(3,1)$ of some energy with representation spaces of other energies.
  It turns out that the totality of all
 states of the atom spans a representation space of $SO(4,2)$,
 which is the maximal
kinematical space-time transformation that leave invariant the
Schrodinger equation for Hydrogen Atom.
 Thus one can say that this dynamic group
encompasses all the atomic states together with their
transformations, including the transformations between states of
different energy \cite {Fano}.  Although there are scattered
results existing in the literature, to the authors's knowledge,
there is no result for both bound and scattering states with
time-dependent Hamiltonians for hydrogen atom. Our study paves a
way for understanding the control of decoherence problem, since
the interaction Hamiltonians as a rule are time-dependent in the
study of the decoherence problem.

In this paper, we begin  in section 2 with summarizing the theorem
of strong analytic controllability of time-dependent quantum
systems\cite{Lan_1}. As the tangent space to $SO(4,2)$ at the
identity, $so(4,2)$ has a good correspondence geometric structure
of the Lie group $SO(4,2)$.  Moreover, $so(4,2)$ is a vector
space, a linear object, so it is easier to study than $SO(4,2)$.
In section 3, we give the realization of $so(4,2)$ for negative
energy and positive energy by solving differential equations,
which is the representation states of $SO(4,2)$ as well. Using
operators from the realization of $so(4,2)$ as interacting
Hamiltonians, we construct a hydrogen atom control system in
section 4.  Based on the above and the strong analytic
controllability theorem\cite{Lan_1}, we conclude that this
hydrogen atom control system is strongly analytically
controllable.  The construction of this hydrogen atom control
system provides theoretical direction on how to construct a new
control system by adding controls and interacting Hamiltonians to
the original Schr\"odinger equation so that this new control
system is strongly analytically controllable. Finally we conclude
this paper with section 5.

\section{A Strong Analytic Controllability Theorem for
Time-dependent Quantum Systems}

Quantum control systems as described by the Schr\"odinger equation
are bilinear systems with respect to controls and states
\begin{equation}
\begin{split}
 \frac{\partial }{\partial t} \psi (t,x) & = (H_0 (t,x) +
\sum_{l=1}^r u_l (t) H_l (t,x)) \psi (t,x), \\
&  \quad \psi (t_0, x) = \psi _0, \label{equ411}
\end{split}
\end{equation}
where $H_0 (t,x), H_l (t,x) $, $l=1,2, \ldots , r$, are
skew-Hermitian operators on unit sphere $S_{\cal H}$, $\psi (t,x)
\in S_{\cal H}$, and $u_l(t), \, l=1, \ldots , r$ are piecewise
constant functions.  In fact, the above skew-Hermitian operator
$H(t,x)$ can be written as $iH^\prime (t,x)$, where $H^\prime $,
being Hermitian, is an observable.

Operators involved in system (\ref{equ411}) are generally
unbounded and act on an infinite-dimensional Hilbert space which
consists of quantum states $\psi $.  In order to find a domain on
which the exponentiations of these operators converge, we
introduce a so-called analytic domain $D_\omega $ proposed by
Nelson \cite {Nelson}, a dense domain invariant under the action
of operators in system (\ref{equ411}).  In addition, the solution
of the Schr\"odinger equation can be expressed globally in
exponential form on this domain, which is also invariant under the
action of the exponentiations of these operators.

For system (\ref{equ411}), a theorem \cite {Huang_1, Huang_2}
tells us that the transitivity of states on $S_{\cal H}$ requires
infinite control manipulations based on the piecewise constant
control set $ \{u_l(t)\}$. In practice, switching control
manipulations infinitely seems meaningless.  So practical interest
directs us to consider controllability on a finite-dimensional
submanifold $M$ of the unit sphere Hilbert space $S_{\cal H}$.
Moreover, a finite-dimensional tangent space generated by $H_0
(t,x) \psi (t,x), \ldots , H_r (t,x) \psi (t,x)$ is defined on $M
\cap D_\omega $ and $\dim M \cap D_\omega =m$ when assuming $\dim
M =m$. That is, the finite-dimensional tangent space generated by
$H_0 (t,x) \psi (t,x), \ldots , H_r (t,x) \psi (t,x)$ is densely
defined on $M$. Further the manifold $M$ can be given by the
closure of the set $\{e^{t_0H_{p0}} e^{t_1H_{p1}} \ldots
e^{t_rH_{pr}}\psi _0 \}$, with $(p0, p1, \ldots , pr)$ any
permutation of  $(0, 1, \ldots , r)$ and $t_i \in R^1,\, i=0, 1,
\ldots , r$.

On an analytic domain, the notion of controllability is modified
by that of analytic controllability.
\begin{definition}
Let $\cal L$ be the Lie algebra generated by skew-Hermitian
operators $H_0, \ldots , H_r$ on a unit sphere $S_{\cal H}$. For
system (\ref{equ411}), if $D_\omega $ exists for $\cal L$, and for
any $\psi _0, \psi _d \in D_\omega \cap M $, there exist $u_1(t),
\ldots , u_r(t)$, and $T$ ($\forall T$) s.t. the solution of
control system (\ref{equ411}) satisfies $\psi (t_0, x) = \psi
_0,\, \psi (T,x) = \psi _d$ and $\psi (t, x) \in  D_\omega \cap M
$, $t_0 \leq t \leq T$, then the system is called analytically
controllable (strongly analytically controllable) on $M$, and we
say  the corresponding unitary Lie group is analytically
transitive on $M$.
\end{definition}
Let $R_t(\psi _0)$ denote the reachable set of system
(\ref{equ411}) from starting state $\psi _0$ at time $t$, and let
$R(\psi _0) = \bigcup _{t>t_0} R_t(\psi _0)$ denote the reachable
set starting from $\psi _0$ at time greater than $t_0$.
\begin{definition}
A time-dependent control system is said to be completely
controllable if $R(\psi _0) = M$ holds for all $\psi _0 \in M$. If
$R_t(\psi _0) = M$ for all time $t>t_0$ and for all $\psi _0 \in
M$, then this system is said to be strongly completely
controllable on $M$.  If $R_t(\psi _0) = M \cap D_\omega $ for all
time $t>t_0$ and for all $\psi _0 \in M \cap D_\omega $, then this
system is said to be strongly analytically controllable on $M$.
\end{definition}

With the help of Kunita' method \cite{Kunita_1}, we derive the
following strong analytic controllability theorem in \cite{Lan_1,
Lan_2}.
\begin{theorem}
Considering quantum control system (\ref{equ411}), let
$$
\begin{array}{l}
{\cal B}(t,x) = {\cal L} ( H_1 (t,x), \ldots , H_r (t,x)) \\
B_1 = -[ H_0, {\cal  B}] + \frac{\partial }{\partial t} {\cal B}\\
\vdots \\
B_n = -[ H_0, B_{n-1} ] + \frac{\partial }{\partial t} B_{n-1} \\
\vdots  \\
{\cal C} = {\cal L} \{{\cal B}, B_1, \ldots , B_n, \ldots \}.
\end{array}
$$
Suppose  $\dim \, {\cal C} (t,x) \psi (t,x) = m$ holds for all
$\psi  \in  M \cap D_\omega $, and $[ {\cal B},  {\cal C} ](t,x)
\subset  {\cal B}(t,x)$ holds for all $(t,x)$. Then the
time-dependent quantum control system (\ref{equ411}) is strongly
analytically controllable. \label{the45}
\end{theorem}

\section{The realization and the representation of $so(4,2)$ }

Kleinert \cite {Kleinert} pointed out that the internal structure
of the quantum mechanical system of the hydrogen atom can be
described completely in terms of simple group operations in the
representation space of the non-compact group $SO(4,2)$. $SO(4,2)$
is the group of rotation in six-dimensional pseudo-Euclidean space
with the metric
\begin{equation}
g = diag (1, 1, 1, 1, -1, -1).
\end{equation}
It is a 15-parameter group and is generated by 15 generators
$L_{\alpha \beta } = - L_{\beta \alpha }$, $\alpha , \beta = 1,
\ldots , 6$ which form the Lie algebra $so(4,2)$ with the
commutation rules
\begin{equation}
[L_{\alpha \beta }, L_{\alpha \gamma }] = i g_{\alpha \alpha }
L_{\beta \gamma }.
\end{equation}
All information contained in the Schr\"odinger theory of the
hydrogen atom can be expressed in a completely algebraic language.
That is, representation states of $so(4,2)$ can be brought in
one-to-one correspondence with the states of the hydrogen atom.

It is known \cite{Adams} that  the Lie algebra $so(4,2)$ satisfies
the following commutation relations:
\begin{align}
&{ [{\bf L}, {\bf L} ]} =  i{\bf L} & & { [{\bf L}, {\bf A} ]} =
i{\bf A}
& & { [{\bf L}, {\bf B} ]} =  i{\bf B} \label{equ524} \\
&{ [{\bf L}, S ]} =  0  & & { [{\bf A}, {\bf A} ]} =  i{\bf L}
& & { [S, {\bf A} ]}  =  i{\bf B} \\
&{ [{\bf A}, {\bf B} ]}  =  i S & & { [{\bf B}, {\bf B} ]}  =
-i{\bf L}
& & { [S, {\bf B} ]}  =  i{\bf A} \\
&{ [C, {\bf A} ]}  =  i{\bf \Gamma } & & { [D, {\bf A} ]}  =  0
& & { [C, {\bf B} ]}  =  0 \\
&{ [D, {\bf B} ]}  =  -i{\bf \Gamma } & & { [C, {\bf L} ]}  =  0 &
& { [D, {\bf L} ]}  =  0 \\
& { [{\bf L}, {\bf \Gamma } ]}  =  i{\bf \Gamma } & &{ [ C, S ]} =
-i D &
&{ [ S,D ]}  =  i C \\
&{ [ D, C ]}  =  i S & &{ [{\bf \Gamma }, {\bf A} ]}  =  -i C &
&{ [{\bf \Gamma }, {\bf B} ]}  =  -i D \\
&{ [{\bf \Gamma }, C ]}  =  -i{\bf A} & &{ [{\bf \Gamma }, S ]}  =
0 &
&{ [{\bf \Gamma }, D ]}  =  -i {\bf B} \\
&{ [{\bf \Gamma }, {\bf \Gamma } ]}  =  -i {\bf L}, \label{equ525}
\end{align}
where $[{\bf A}, {\bf B} ] = iS$ means $[A_i, B_j ] = i \delta
_{ij} S$.

A realization of the Lie algebra $so(4,2)$ is a homomorphism which
associates a concrete set of operators with abstract basis vectors
of the Lie algebra \cite{Adams}.  It is more of a physical concept
than a mathematical one. In quantum mechanics, such operators are
often differential operators expressed in terms of the position
operator ${\bf r}$ and momentum operator ${\bf p}$, which act on a
Hilbert space  of quantum mechanical states. In the following we
will give a realization of $so(4,2)$ for negative energy and
positive energy respectively by solving differential equations
based on Bacry's method, with which Bacry derived the realization
of $so(4,1)$ for negative energy \cite {Bacry}. Dothan \cite
{Dothan} proposed that these generators constitute a
finite-dimensional spectrum generating algebra. Adams \cite
{Adams} described a beautiful algebraic method to construct
another type of 15 generators of $so(4,2)$. These two sets of 15
generators are equivalent according to the $so(4,2)$ commutating
structure.

For the negative energy case, Greiner  and M\"uller \cite
{Greiner_1} describes a way to relate $so(4)$  to the hydrogen
atom Hamiltonian
\begin{equation}
H= \frac{p^2}{2} - \frac{1}{r}, \label{equ528}
\end{equation}
expressed in the atomic units, where $p$ is the momentum and $r$
is the distance between the electron and the proton in the
hydrogen atom. That is, they showed that the angular momentum $
{\bf L} = {\bf r} \times {\bf p}$ and the Runge-Lenz vector ${\bf
A} = \frac{1}{\sqrt {-2H}}{\frac{1}{2} ({\bf L} \times {\bf p} -
{\bf p} \times {\bf L}) + \frac {\bf r}{r}}$ generate the Lie
algebra $so(4)$, which is part of the Lie algebra of $so(4,2)$.
Moreover, ${\bf L}$ and ${\bf A}$ commute with Hamiltonian $H$.
Hence, these operators can only transform a quantum state to
another with the same energy level. When quantum systems interact
with the environment, they may absorb or release energy, thus
changing their energy levels.  $so(4,2)$ is more interesting
because it permits to transform an energy level into another
energy level \cite {Adams, Dothan, Kleinert, Barut, Barut_1}.
Hence, $so(4,2)$ is represented fully on the energy eigenstates,
 be they discrete or continuous, and energy degenerate or
nondegenerate.

In order to find the other operators ${\bf B}, {\bf \Gamma }, S,
C$, and $D$ in terms of ${\bf r}$ and ${\bf p}$, it is much
simpler to use the Poisson bracket instead of commutators because
equations (\ref{equ524}) --(\ref{equ525}) lead to differential
equations which can be integrated in a simple way under the
Poisson bracket \cite{Bacry}.  Let $f({\bf r}, {\bf p})$ be a
scalar function of the classical variables ${\bf r}$ and ${\bf
p}$, satisfying the Poisson bracket relation $\{ {\bf L}, f \}=0$.
$f$ can be written as a function of three scalars $r$, $H$, and
$l$, where $l^2 = r^2 p^2 - ( {\bf r} \cdot {\bf
p})^2$\cite{Bacry}. Then we solve the differential equations in
the Poisson bracket formulism. The operators so calculated in
terms of the Poisson bracket may need be symmetrized to be
Hermitian to get the following negative-energy realization of
$so(4,2)$ \cite{Lan_2}.
\begin{align}
& {\bf L} = {\bf r} \times {\bf p} \label {equ5344}\\
& {\bf A} =  \frac{1}{\sqrt {-2H}} [ ( {\bf L} \times {\bf p} -
{\bf
p} \times {\bf L})/2 - \frac{{\bf r}}{r} ] \\
& {\bf B}  =   \frac{{\bf p}r+r{\bf p}}{2}\cos \zeta -
\frac{1}{\sqrt {-2H}} \frac{{\bf p}({\bf r} \cdot {\bf p})+ ({\bf
p} \cdot {\bf r}){\bf p}}{2} \sin \zeta  \nonumber  \\
& \quad \quad + \frac{1}{\sqrt {-2H}}\frac{{\bf r}}{r} \sin \zeta
\\
& {\bf \Gamma }  =  -\frac {{\bf p}r+r{\bf p}}{2} \sin \zeta -
\frac{1}{\sqrt {-2H}}\frac{{\bf p}({\bf r} \cdot {\bf p})+({\bf p}
\cdot {\bf r}){\bf p}}{2} \cos \zeta  \nonumber  \\
& \quad \quad + \frac{1}{\sqrt
{-2H}}\frac{{\bf r}}{r} \cos \zeta \\
& S  =  - \frac{{\bf r} \cdot {\bf p}+{\bf p} \cdot {\bf r}}{2}
\sin \zeta - \frac{1}{\sqrt
{-2H}}(1+2Hr) \cos \zeta  \\
& C  =   - \frac{{\bf r} \cdot {\bf p}+{\bf p}\cdot {\bf r}}{2}
\cos \zeta + \frac{1}{\sqrt
{-2H}}(1+2Hr) \sin \zeta \\
& D =  \frac{1}{\sqrt {-2H}}, \label {equ5345}
\end{align}
where $ \zeta = \sqrt {-2H}({\bf r} \cdot {\bf p}+{\bf p} \cdot
{\bf r})/{2} + (-2H)^{\frac{3}{2}}t$.

Further, we compute the commutators between the Hamiltonian $H$
and the generators of $so(4,2)$ as follows.
\begin{align}
& {[ H, {\bf L} ] } =  0 \label {equ5346} \\
&{[ H, {\bf A} ]}  =  0 \\
&{[ H, {\bf B} ]}  =  (-2H)^{\frac{3}{2}} i {\bf \Gamma } \\
&{[ H, {\bf \Gamma } ]}  =  - (-2H)^\frac{3}{2} i {\bf B} \\
&{[ H, S ]}  =  (-2H)^\frac{3}{2} i C \\
&{[ H, C ]}  =  -(-2H)^\frac{3}{2} i S \\
&{[ H, D ]}  =  0.  \label {equ5347}
\end{align}
Obviously these $so(4,2)$ generators satisfy the condition
\begin{equation}
i\frac{\partial G}{\partial t} - [H, G] =0.
 \label{equ5348}
\end{equation}
In \cite {Dothan}, the relation (\ref{equ5348}) tells us that if
$\psi $ is an eigenstate of the Schr\"odinger equation
\begin{equation}
i \frac{\partial \psi }{\partial t} = H \psi , \label {equ5349}
\end{equation}
then $G \psi $ is an eigenstate of the above Schr\"odinger
equation when $G$ satisfies the condition (\ref{equ5348}).
 So this realization of $so(4,2)$ forms a
finite-dimensional spectrum-generating algebra \cite {Dothan}. All
of the energy eigenfunctions of the physical problem form a basis
for a single unitary irreducible representation.

\noindent {\it Remark 3.1   } \quad We know that $D$, ${\bf L}$,
and ${\bf A}$ commute with the Hamiltonian $H$, so the multiplets
produced by the action of the transformations generated by $D$,
${\bf L}$, and ${\bf A}$ are degenerate eigenfunctions of $H$. The
other generators of $so(4,2)$ do not commute with Hamiltonian $H$,
so the multiplets produced by the transformation generated by
these operators cannot be degenerate eigenfunctions of $H$.
Rather, these transformations generate a spectrum of
eigenfunctions of $H$.

For the positive energy, it is known \cite{Adams} that the angular
momentum $ {\bf L} = {\bf r} \times {\bf p}$ and the Runge-Lenz
vector ${\bf B} = \frac{1}{\sqrt {2H}}{\frac{1}{2} ({\bf L} \times
{\bf p} - {\bf p} \times {\bf L}) + \frac {\bf r}{r}}$ generate
the Lie algebra $so(3,1)$, which is part of the Lie algebra of
$so(4,2)$.  Similar to the negative energy case, we give  the
positive-energy realization of $so(4,2)$ as follows.
\begin{align}
& {\bf L}= {\bf r} \times {\bf p} \label {equ5416}\\
& {\bf A} =  \frac{{\bf p}r+r{\bf p}}{2}\sinh \zeta -
\frac{1}{\sqrt {2H}}\frac{{\bf p}({\bf r} \cdot {\bf p})+({\bf p}
\cdot {\bf r}){\bf p}}{2} \cosh \zeta \nonumber \\
& \quad \quad + \frac{1}{\sqrt {2H}}\frac{{\bf r}}{r}\cosh \zeta  \\
& {\bf B}  =  \frac{1}{\sqrt {2H}} [ ( {\bf L} \times {\bf p} -
{\bf
p} \times {\bf L})/2 - \frac{{\bf r}}{r} ] \\
& {\bf \Gamma }  =   \frac{{\bf p} r +r{\bf p}}{2}\cosh \zeta -
\frac{1}{\sqrt {2H}}\frac{{\bf p}({\bf r} \cdot {\bf p})+({\bf p}
\cdot {\bf r}){\bf p}}{2} \sinh \zeta  \nonumber \\
& \quad \quad +\frac{1}{\sqrt {2H}}
\frac{{\bf r}}{r} \sinh \zeta \\
& S =  \frac{1}{\sqrt
{2H}} (2Hr+1) \sinh \zeta- \frac{{\bf r} \cdot {\bf p}+{\bf p} \cdot {\bf r}}{2} \cosh \zeta \\
& D =  \frac{{\bf r} \cdot {\bf p}+{\bf p} \cdot {\bf r}}{2} \sinh
\zeta -\frac{1}{\sqrt
{2H}} (2Hr+1) \cosh \zeta \\
& C = \frac{1}{\sqrt {2H}}, \label {equ5417}
\end{align}
where $ \zeta = \sqrt {2H}({\bf r} \cdot {\bf p}+{\bf p} \cdot
{\bf r})/2 - (2H)^{\frac{3}{2}}t$.

Similar to the negative energy case, we compute the commutation
relations between the Hamiltonian $H$ and these fifteen
positive-energy generators of $so(4,2)$ as follows.
\begin{eqnarray}
{[ H, {\bf L} ] }& = & 0 \label {equ5418}  \\
{[ H, {\bf A} ]} & = & -(2H)^\frac{3}{2} i {\bf \Gamma } \\
{[ H, {\bf B} ]} & = & 0 \\
{[ H, {\bf \Gamma } ]} & = & - (2H)^\frac{3}{2} i {\bf A} \\
{[ H, S ]} & = & (2H)^\frac{3}{2} i D \\
{[ H, D ]} & = & (2H)^\frac{3}{2} i S \\
{[ H, C ]} & = & 0.  \label {equ5419}
\end{eqnarray}
It is easily verified that the realization of $so(4,2)$ for
positive energy satisfies the relationship (\ref{equ5348}), based
on the relationships (\ref{equ5418}) --- (\ref{equ5419}).

\noindent {\it Remark 3.2   } \quad In fact, relation
(\ref{equ5348}) is a special case of the following relation.  Let
$Q = i\partial _t -H,$ where $H$ is the Hamiltonian. Then the
sufficient condition for operator $Z$ to belong to a symmetry
algebra is \cite{Miller_3}
\begin{equation}
[Z, Q] = R_L(t,x)Q, \label{equ5348_new}
\end{equation}
where $R_L(t,x)$ is a operator having consistent orders of
$\partial t$ and $\partial x$, and $x \in R^n$.  When
$R_L(t,x)=0$, relation (\ref{equ5348_new}) is reduced  to relation
(\ref{equ5348}).  In \cite{Miller_3}, Miller uses relation
(\ref{equ5348_new}) to find the basis of the symmetry algebra,
then diagonalize different operators in this symmetry algebra to
separate variables in different coordinate systems, thereby
finding solutions for differential equations.

Now let us study the matrix representation of $\cal L$.  If $\{
|n\rangle  : n=1, \ldots , N \}$, where $N$ can be infinity, is a
basis for vector space ${\cal V}$, and if $P_i=T(h_i), i =1,
\ldots, l$, where $h_i \in {\cal L}$, $h_i$ is a basis for ${\cal
L}$, and $T$ is a homomorphism, then $P_i|n\rangle  =\sum_m
|m\rangle \langle m|P_i|n\rangle ,$ $ m, n = 1, \ldots , N$, where
$\langle m|P_i|n\rangle  $ denotes matrix element $(m,n)$ of
$P_i$. These matrices are a basis for a matrix representation of
the Lie algebra $\cal L$.  The above realizations of $so(4,2)$ can
now be employed for us to look for the representation states $\{
|n\rangle  : n=1, \ldots , N \}$. First, we decompose $so(4,2)$ as
follows
$$
so(4,2) \supset so(4) \otimes so(2,1)
$$
Following the method in \cite{Adams} of expanding the
representation states of $so(4)$  and the representation states of
$so(2,1)$, we obtain the representation states of $so(4,2)$ for
negative energy case as follows:
\begin{align}
& L_3 |nlm\rangle   =  m|nlm\rangle  \label {equ554} \\
& L_+ |nlm\rangle   =  \omega _m^l |nl,m+1\rangle  \\
& L_- |nlm\rangle   =  \omega _{-m}^l |nl,m-1\rangle \\
& A_3 |nlm\rangle   =  \alpha _m^l c_l^n |n,l-1,m\rangle  + \alpha
_m^{l+1}
c_{l+1}^n |n,l+1,m\rangle  \\
& A_+ |nlm\rangle   =  \beta _m^{l-1}c_l^n |n,l-1,m+1\rangle \nonumber \\
& \quad  - \gamma
_m^{l+1}c_{l+1}^n |n, l+1, m+1\rangle  \\
& A_- |nlm\rangle  =  -\beta _{-m}^{l-1} c_l^n |n,l-1, m-1\rangle \nonumber \\
& \quad  + \gamma
_{-m}^{l+1}c_{l+1}^n |n,l+1,m-1\rangle \\
& B_3 |nlm\rangle   = \alpha _m^l u_l^n |n-1, l-1,m\rangle
 + \alpha _m^l v_l^n \nonumber \\
 & \quad    \cdot |n+1,l-1,m\rangle + \alpha _m^{l+1} v_{l+1}^{n-1} |n-1,l+1,m\rangle  \nonumber \\
&  \quad   + \alpha _m^{l+1}
u_{l+1}^{n+1} |n+1,l+1,m\rangle  \\
& B_+ |nlm\rangle  = \beta _m^{l-1} u_l^n |n-1,l-1,m+1\rangle  +
\beta _m^{l-1}
v_l^n  \nonumber \\
& \quad \cdot |n+1,l-1,m+1\rangle  - \gamma _m^{l+1}
v_{l+1}^{n-1} |n-1, l+1,m+1\rangle \nonumber \\
& \quad  - \gamma _m^{l+1} u_{l+1}^{n+1} |n+1, l+1, m+1\rangle  \\
& B_- |nlm\rangle   = -\beta _{-m}^{l-1} u_l^n |n-1,l-1,m-1\rangle
- \beta
_{-m}^{l-1} v_l^n \nonumber \\
& \quad  \cdot |n+1,l-1,m-1\rangle   + \gamma _{-m}^{l+1}
v_{l+1}^{n-1} |n-1, l+1,m-1\rangle \nonumber \\
& \quad +\gamma _{-m}^{l+1} u_{l+1}^{n+1} |n+1, l+1, m-1\rangle \\
& \Gamma _3 |nlm\rangle  = -i\alpha _m^l u_l^n |n-1, l-1,m\rangle
+ i\alpha
_m^l v_l^n \nonumber \\
& \quad  \cdot |n+1,l-1,m\rangle
 -i\alpha _m^{l+1} v_{l+1}^{n-1} |n-1,l+1,m\rangle \nonumber \\
& \quad  + i\alpha _m^{l+1} u_{l+1}^{n+1} |n+1,l+1,m\rangle  \\
& \Gamma _+ |nlm\rangle   = -i\beta _m^{l-1} u_l^n |n-1,
l-1,m+1\rangle  + i\beta
_m^{l-1} v_l^n \nonumber \\
& \quad  \cdot |n+1,l-1,m+1\rangle  +i\gamma _m^{l+1}
v_{l+1}^{n-1} |n-1,l+1,m+1\rangle \nonumber \\
& \quad  - i\gamma _m^{l+1} u_{l+1}^{n+1} |n+1,l+1,m+1\rangle
\end{align}
\begin{align}
& \Gamma _- |nlm\rangle   = i\beta _{-m}^{l-1} u_l^n |n-1,
l-1,m-1\rangle -
i\beta _{-m}^{l-1} v_l^n \nonumber \\
& \quad \cdot |n+1,l-1,m-1\rangle  -i\gamma _{-m}^{l+1}
v_{l+1}^{n-1} |n-1,l+1,m-1\rangle \nonumber \\
& \quad + i\gamma _{-m}^{l+1} u_{l+1}^{n+1} |n+1,l+1,m-1\rangle  \\
& D |nlm\rangle   = n |nlm\rangle \\
& T_+ |nlm\rangle  = \omega _l^n |n+1,lm\rangle  \\
& T_- |nlm\rangle  = \omega _l^{-n} |n-1,lm\rangle ,
 \label{equ555}
\end{align}
where
\begin{align*}
& \alpha _m^l  =  \sqrt {(l-m)(l+m)} , & u_l^n = \frac{1}{2} \sqrt
{\frac{(n+l-1)(n+l)}{(2l-1)(2l+1)}},
\end{align*}
\begin{align*}
& c_l^n = \sqrt {\frac{(n-l)(n+l)}{(2l-1)(2l+1)}}, & v_l^n =
\frac{1}{2} \sqrt {\frac{(n-l)(n-l+1)}{(2l-1)(2l+1)}},
\end{align*}
and $m= -l, \ldots ,l-1, l,\quad l= 0,1, \ldots , n-1,\quad n=
1,2,3, \ldots $.

\noindent {\it Remark 3.3  } \quad Vector operators ${\bf L}$ and
${\bf A}$ act on the angular portion of the basis kets
$|nlm\rangle $, affecting only the magnetic quantum number $m$ and
orbital angular momentum quantum number $l$, while operators $S,
C$, and $D$ act only on the remaining part. Further, vector
operators ${\bf B}$ and ${\bf \Gamma }$ act on both parts.

Construction of the representation of $so(4,2)$ for positive
energy follows a procedure similar to that used above for the
negative energy case,  based on the following alternative
decomposition
$$
so(4,2) \supset so(3,1) \otimes so(2,1).
$$
For the radial part $S, C, D$ of this realization
 we need to diagonalize $C$, which has a continuous spectrum,
 instead of diagonalizing  $D$ in the negative energy case.
 $$
C |vlm\rangle  = v |vlm\rangle , \quad v \in (-\infty,  \infty),
$$
where the $|vlm\rangle $ denote the continuous eigenstates. Thus,
using $|vlm\rangle $ to denote the continuous eigenstates, the
representation of $so(4,2)$ for positive energy is similar to that
for negative energy with the following exceptions --- replacing
quantum number $n$ in the negative energy case with $iv$, $l=0, 1,
\ldots , \infty $, and  $v \in (-\infty , \infty )$ being
continuous \cite {Barut_1, Kleinert}.

\section{ A strongly and analytically controllable hydrogen atom system }

Now let us consider the hydrogen atom control system
\begin{equation}
\begin{split}
& i\frac{\partial }{\partial t} \psi (t,x) =  [ H+u_1(t)L_1 +
u_2(t)L_2 +u_3(t)L_3   \\
& \quad +u_4(t)A_1+u_5(t)A_2 +u_6(t)A_3  +u_7(t)B_1  \\
& \quad  +u_8(t)B_2 +u_9(t)B_3 +u_{10}(t)\Gamma _1+u_{11}(t)
\Gamma _2   \\
& \quad  +u_{12}(t)\Gamma _3+u_{13}(t) S +u_{14}(t)C+ u_{15}(t)D ]
\psi (t,x), \label {equ561}
\end{split}
\end{equation}
where $H,L_i, A_i, B_i, \Gamma _i, S,C,D, i=1,2,3$ are Hermitian
operators defined as before, and $u_j(t), j=1,2,\ldots , 15$ are
piecewise constant control functions. Divide the operators in the
right hand side of equation (\ref{equ561}) by $i$, converting  the
control system (\ref{equ561}) to the standard form
\begin{equation}
\begin{split}
& \frac{\partial }{\partial t} \psi (t,x)  =  [ H^\prime
+u_1(t)L_1^\prime + u_2(t)L_2^\prime +u_3(t)L_3^\prime  \\
& \, +u_4(t)A_1^\prime +u_5(t)A_2^\prime +u_6(t)A_3^\prime
+u_7(t)B_1^\prime  \\
& \, +u_8(t)B_2^\prime +u_9(t)B_3^\prime +u_{10}(t)\Gamma
_1^\prime +u_{11}(t)
\Gamma _2^\prime  \\
& \, +u_{12}(t)\Gamma _3^\prime   +u_{13}(t) S^\prime
+u_{14}(t)C^\prime + u_{15}(t)D^\prime ] \psi (t,x), \label
{equ562}
\end{split}
\end{equation}
where $H^\prime ,L_i^\prime , A_i^\prime , B_i^\prime , \Gamma
_i^\prime , S^\prime ,C^\prime ,D^\prime , i=1,2,3$ are
skew-Hermitian operators, and the $u_j(t)$, $ j=1,2,\ldots , 15$,
are piecewise constant functions. Corresponding to the commutators
(\ref{equ524})--(\ref{equ525}), we have
\begin{align}
&[{\bf L^\prime }, {\bf L^\prime } ]  =  {\bf L^\prime } & &[{\bf
L^\prime }, {\bf A^\prime } ]  =  {\bf A^\prime } &
&[{\bf L^\prime }, {\bf B^\prime } ]  =  {\bf B^\prime }  \label {equ563} \\
&[{\bf L^\prime }, S]  =  0  & &[{\bf A^\prime }, {\bf A^\prime }
]  =  {\bf L^\prime } &
&[S, {\bf A^\prime } ]  =  {\bf B^\prime } \\
&[{\bf A^\prime }, {\bf B^\prime } ]  =  S^\prime & &[{\bf
B^\prime }, {\bf B^\prime } ]  =  -{\bf L^\prime } &
&[S^\prime , {\bf B^\prime } ]  =  {\bf A^\prime } \\
&[C^\prime , {\bf A^\prime } ]  =  {\bf \Gamma ^\prime } &
&[D^\prime , {\bf A^\prime } ]  =  0 &
&[C^\prime , {\bf B^\prime } ]  =  0 \\
&[D^\prime , {\bf B^\prime } ]  =  -{\bf \Gamma ^\prime } &
&[C^\prime , {\bf L^\prime } ]  =  0 &
&[D^\prime , {\bf L^\prime } ]  =  0 \\
&[{\bf L^\prime }, {\bf \Gamma ^\prime } ]  =  {\bf \Gamma ^\prime
} & &[C^\prime , S^\prime ]  =  -D^\prime &
&[S^\prime , D^\prime ]  =  C^\prime \\
&[D^\prime , C^\prime ]  =  S^\prime & &[{\bf \Gamma ^\prime },
{\bf A^\prime } ]  =  -C^\prime  &
&[{\bf \Gamma ^\prime }, {\bf B^\prime } ]  =  -D^\prime  \\
&[{\bf \Gamma ^\prime }, C^\prime  ]  =  -{\bf A^\prime } & &[{\bf
\Gamma ^\prime }, S^\prime  ]  =  0 &
&[{\bf \Gamma ^\prime }, D^\prime  ]  =  -{\bf B^\prime } \\
&[{\bf \Gamma ^\prime }, {\bf \Gamma ^\prime } ]  =  -{\bf
L^\prime }. \label {equ564}
\end{align}
From the commutators (\ref{equ563})--(\ref{equ564}), we can see
that ${\bf L^\prime }, {\bf A^\prime }, {\bf B^\prime }, {\bf
\Gamma ^\prime }, S^\prime , C^\prime , D^\prime $ form a real Lie
algebra, which we denote by $\overline {\cal B}$.  In addition, it
has already been shown that both of the realizations of $so(4,2)$
for positive and negative energy satisfy relation (\ref{equ5348}).
So we can rewrite relation (\ref{equ5348}) as
\begin{equation}
\frac{\partial G^\prime }{\partial t} - [ H^\prime , G^\prime ]
=0,\quad \mbox{ where } G^\prime  \in \overline {\cal B}.
\end{equation}
Based on the strong analytic controllability theorem given in
section 2, the following holds:
$$B_1 ={\cal L}\{-[H^\prime
, G^\prime ] + \frac{\partial G^\prime }{\partial t}, \quad
G^\prime \in \overline {\cal B} \} = 0,$$ so $\overline {\cal C}
={\cal L}\{\overline {\cal B}, B_1, \ldots \} = \overline {\cal
B}$ and $[\overline {\cal B}, \overline {\cal C} ] \subset
\overline {\cal B}$ . In addition, the representation space of
$so(4,2)$ spanned by $|nlm\rangle  $ and $|vlm\rangle  $ for
negative and positive energies respectively, is indeed an analytic
domain, because this space is invariant both under the Lie algebra
$so(4,2)$ and under the Lie group $SO(4,2)$.  That is,
exponentiation of any operator from $so(4,2)$ acting on this
representation space converges. Many examples can be found in
\cite {Miller_2}.   Let $M$ denote the closure of a manifold
spanned by a finite set of eigenfunctions of $so(4,2)$ as is the
case in \cite{Huang_1, Huang_2, Clark_1, Clark_2}. It is not hard
to see that $\dim \overline {\cal C} \psi (x,t) = \dim M$
\cite{Huang_1, Huang_2, Clark_1, Clark_2} for any $(x,t)$
according to the representation of $so(4,2)$ in the last section.
Hence, by Theorem \ref{the45}, the control system (\ref{equ561})
is strongly analytically controllable on $M$. In fact, the effect
of the drift term $H\psi $ on the control system can  be
compensated by the effect of $\overline {\cal B} \psi $.

\noindent {\it Remark 4.1   } \quad In control system
(\ref{equ562}), we do not need 15 generators to appear in the
control system.  We can selectively pick up at least five
generators, for example, $L_1^\prime , L_2^\prime , A_3^\prime ,
S^\prime $, and $C^\prime $, and then by the commutation relations
(\ref{equ563})--(\ref{equ564}) we can generate the other
generators so as to make the system (\ref{equ562})  strongly
analytically controllable as 15 generators do.   Because system
(\ref{equ562}) corresponds to system (\ref{equ561}), we
immediately conclude that the system
\begin{align}
i\frac{\partial }{\partial t} \psi (t,x)  = & [H+u_1(t)L_1
+u_2(t)L_2
+u_3(t) A_3 \nonumber \\
& + u_4(t)S + u_5(t)C] \psi (t,x) \label{equ565}
\end{align}
is strongly analytically controllable.

The detailed analysis presented in this section provides
theoretical guidance for choosing and imposing controls and
interacting Hamiltonians on an original unperturbed system
described by the Schr\"odinger equation for the hydrogen atom,
such that the constructed control system is strongly analytically
controllable. In addition, we can see that the unperturbed
Hamiltonian (\ref{equ528}) and the interacting Hamiltonians from
the realization of $so(4,2)$ act on the same state space. That is,
addition of interacting Hamiltonians does not affect the state
space of the original hydrogen atom system. The controllability
problem can still be analyzed on the original state space, but
with imposing controls and the interacting Hamiltonians.

Note that the angular momentum and the Runge-Lenz vector commute
with the hydrogen atom Hamiltonian, so they cannot cause a
transition from one energy level quantum state to another.
However, the rest of operators of the realization of $so(4,2)$ can
produce such transitions. Hence, at least one operator besides the
angular momentum and the Runge-Lenz vector must be chosen as
interacting Hamiltonians (as we did in control systems
(\ref{equ561}) and (\ref{equ565})) in order that the system is
strongly analytically controllable among different quantum states
of different energies.

\section{Conclusion}

Studying the manipulation of hydrogen atom --- the simplest atom
--- can help us understand the manipulation of the other atom
systems. It is known that representation states of $so(4, 2)$ can
be brought in one-to-one correspondence with the discrete,
continuous, degenerate and non-degenerate states of the hydrogen
atom \cite {Adams, Dothan}. In this paper, we first have given the
realization and representation of the Lie algebra of $so(4, 2)$.
Then picking up operators from the realization of $so(4, 2)$ as
interacting Hamiltonians, and choosing piecewise constant
functions as controls, we have constructed an ideal hydrogen atom
control system which is strongly analytically controllable. When
the system is interacting with environment, one has to consider
adding disturbing noises. The results of this paper are the strong
analytic controllability for hydrogen atom control systems without
considering design schemes for control inputs. Based on results of
this paper, we are investigating optimal transitions from bound
states to bound states, scattering states to scattering states,
bound states to scattering states and vice versa. This will
facilitate the selection of desired control inputs to the system.

\section*{Acknowledgment}

This research was supported in part by the U.~S.\ Army Research
Office (TJT) under Grant W911NF-04-1-0386 and by the U.~S.\
National Science Foundation under Grants DMS01-03838 (QSC) and
PHY-0140316 (JWC).  JWC would also like to acknowledge partial
support from FCT POCTI, FEDER in Portugal and the hospitality of
the Centro de Ci\^encias Mathem\'aticas at the Madeira Math
Encounters.

\end{document}